\UseRawInputEncoding
\documentclass[journal]{IEEEtran}
\usepackage[inline]{enumitem}
\usepackage{cite}
\usepackage{amsmath}
\usepackage{amsfonts}
\usepackage{multirow}
\usepackage{subfigure}
\usepackage{graphicx}
\usepackage{algorithmic}
\usepackage{array}
\usepackage{booktabs}
\usepackage[mathscr]{euscript}
\usepackage[hyphens]{url}
\usepackage{makecell}
\usepackage{tablefootnote}
\usepackage{diagbox}
\usepackage{hhline}
\usepackage{pifont}

\newcommand{\cmark}{\ding{51}}%
\newcommand{\xmark}{\ding{55}}%

\usepackage[]{review}
\setrevision{2}

\begin{document}
\title{SpatialNet: Extensively Learning Spatial Information for Multichannel Joint Speech Separation, Denoising and Dereverberation}

\author{Changsheng~Quan, 
        ~Xiaofei~Li
\thanks{Changsheng Quan is with Zhejiang University and Westlake University \& Westlake Institute for Advanced Study, Hangzhou, China. e-mail: quanchangsheng@westlake.edu.cn}
\thanks{Xiaofei Li is with the School of Engineering, Westlake University, Hangzhou 310030, China, and also with the Institute of Advanced Technology, Westlake Institute for Advanced Study, Hangzhou 310024, China. Corresponding author: lixiaofei@westlake.edu.cn. }
}

%

\setlength{\doublerulesep}{0pt}

\maketitle

\begin{abstract}
This work proposes a neural network to extensively exploit spatial information for multichannel joint speech separation, denoising and dereverberation, named SpatialNet.  
In the short-time Fourier transform (STFT) domain, the proposed network performs end-to-end speech enhancement. It is mainly composed of interleaved narrow-band and cross-band blocks to respectively exploit narrow-band and cross-band spatial information. 
The narrow-band blocks process frequencies independently, and use self-attention mechanism and temporal convolutional layers to respectively perform spatial-feature-based speaker clustering and temporal smoothing/filtering. 
The cross-band blocks process frames independently, and use full-band linear layer and frequency convolutional layers to respectively learn the correlation between all frequencies and adjacent frequencies.
Experiments are conducted on various simulated and real datasets, and the results show that 1) the proposed network achieves the state-of-the-art performance on almost all tasks; 2) the proposed network suffers little from the spectral generalization problem; and 3) the proposed network is indeed performing speaker clustering (demonstrated by attention maps). 
\end{abstract}


%
\IEEEpeerreviewmaketitle

\section{Introduction}


Microphone array signal processing is widely used in many applications, like hearing aid, robot and smart home equipment (e.g. smart speaker).
In real-life applications, the speech recordings are inevitably impaired by ambient noise, room reverberation, and interfering speech signals.
Suppressing these interferences, namely removing noises and reverberations, separating speakers, from the recorded signals can improve the speech quality and promote the accuracy of automatic speech recognition (ASR).
This work aims to design a neural network for jointly performing multichannel speech separation, denoising and dereverberation. In the following, for presentation simplicity, sometimes we use the term `speech enhancement' indiscriminately to refer to either one of the three tasks or the joint task.  

In multichannel recordings, there are two kinds of information can be leveraged for speech enhancement, namely spectral information and spatial information \cite{vincent_BlindGuidedAudio_2014, gannot_consolidated_2017}, and each of them has their respective temporal dynamics and dependencies. Spectral information mainly refers to the spectral components (spectral pattern) of signals, which carries the signal content. Spatial information mainly refers to the information of signal propagation and sound field. Traditional speech enhancement methods mainly leverage the spatial information, and thus normally agnostic to the signal's spectral content. Speech processing is normally performed in time-frequency (T-F) domain, by applying short-time Fourier transform (STFT), and speech enhancement methods are formulated in narrow-band. 
Beamforming (spatial filtering), e.g. the minimum variance/power distortionless response (MVDR/MPDR) beamformer, is one fundamental speech enhancement method \cite{Markovich-Golan2018b,gannot2001signal,gannot_consolidated_2017}, which suppresses noise and undesired speakers by applying linear spatial filtering onto the microphone signals. Weighted prediction error (WPE) \cite{nakatani2010speech} is one popular speech dereverberation technique. Based on the narrow-band convolutive transfer function (CTF) model \cite{talmon2009convolutive,li2019multichannel}, WPE applies linear prediction to perform inverse filtering onto the microphone signals. 
In recent years, beamforming and WPE are combined for simultaneous denoising/separation and dereverberation, such as the Weighted Power minimization Distortionless response (WPD) technique \cite{nakatani_UnifiedConvolutionalBeamformer_2019} combines MPDR with WPE. 
Spatial clustering is one popular blind source separation method, where T-F bins are clustered according to the spatial vectors (or spatial cues like inter-channel phase/level difference, IPD/ILD) of different speakers \cite{winter_map-based_2006,boeddecker_front-end_2018}. According to the W-disjoint orthogonality assumption \cite{yilmaz_blind_2004}, each T-F bin is considered to be dominated by a single speaker. The T-F bins belong to the same speaker would have identical/correlated spatial vectors (when the speaker is static), and thus can be clustered together.

There are several ways to leverage deep neural networks for multichannel speech enhancement. 
One way is to perform end-to-end multichannel speech enhancement, such as FasNet with transform-average-concatenate (TAC) \cite{luo_end--end_2020}, MC-TasNet \cite{gu_EndtoEndMultiChannelSpeech_2019}, Channel-Attention (CA) Dense U-net \cite{tolooshams_ChannelAttentionDenseUNet_2020}, etc.
Another important technique is the so-called neural beamformer, which estimates the spatial filter using the neural enhanced (multichannel) signals/masks \cite{heymann_blstm_2015, ochiai_beam-tasnet_2020,wang_MultimicrophoneComplexSpectral_2021, chen_BeamGuidedTasNetIterative_2022,  gu_UnifiedAllNeuralBeamforming_2023}. In the two-stage methods, such as \cite{wang_MultimicrophoneComplexSpectral_2021,chen_BeamGuidedTasNetIterative_2022}, the neural beamformer is followed by a neural post-processing. Neural beamformer is currently the main stream research direction, mainly due to that the linear spatial filtering of beamforming is friendly to the ASR backend.

As discussed above, spatial information are well-formulated in narrow-band, in the form of steering vector, covariance matrix, IPD/ILD, CTF, etc.
Accordingly, traditional methods like beamforming, WPE and spatial clustering are all performed in narrow-band.
Besides, many other important signal properties are also formulated in narrow-band, for example the signal stationarity \cite{gannot2001signal} and spatial coherence \cite{gannot_consolidated_2017,habets2008generating,mohan2008localization}, which are important for discriminating between speech and noise. 
Inspired by these facts, our previous works proposed to leverage the narrow-band information especially the narrow-band spatial information using a neural network for multichannel speech denoising \cite{li2019waspaa,li_narrow-band_2019} and separation \cite{quan_MultiChannelNarrowBandDeep_2022,quan_MultichannelSpeechSeparation_2022}. The same narrow-band network is used to process each STFT frequency independently. \cite{li2019waspaa,li_narrow-band_2019,quan_MultiChannelNarrowBandDeep_2022} all use a simple two-layer long short-term memory (LSTM) network, and achieve promising speech denoising and separation performance, which demonstrates that the narrow-band network is indeed suitable for denoising and separation. In addition, \cite{quan_MultiChannelNarrowBandDeep_2022} shows that the narrow-band network works well for reverberant speech, which means it also suitable for modelling/processing reverberation. 
\addnote[disc1]{1}{Similar discussions about the importance of leveraging narrow-band information can also be found in \cite{wang_TFGridNetIntegratingFull_2022}.}

\textbf{Proposed Method}. As a continuation work of our previous narrow-band networks \cite{li2019waspaa,li_narrow-band_2019,quan_MultiChannelNarrowBandDeep_2022,quan_MultichannelSpeechSeparation_2022}, this work proposes a more powerful network architecture to extensively exploit spatial information, for jointly performing multichannel speech separation, denoising and dereverberation, and it is named SpatialNet. 
The proposed network is composed of a narrow-band block and a cross-band block. 
The narrow-band block is a revision of our previously proposed narrow-band conformer (NBC) network \cite{quan_MultichannelSpeechSeparation_2022}. 
The narrow-band block processes each frequency independently, and is shared by all frequencies. It consists of a multi-head self-attention (MHSA) module and a time-convolutional module. 
One important functional of the MHSA module is to cluster the spatial vector/feature of different frames dominated by different speakers, as is done in \cite{winter_map-based_2006, boeddecker_front-end_2018}. Clustering spatial vectors shares a similar principle with the self-attention mechanism \cite{vaswani2017attention} in the sense of computing the similarity of vectors and then aggregating similar vectors. 
The time-convolutional module is designed for performing local smoothing on the signals, and for modelling the convolutional reverberation. 
Speech and noise signals are random processes, and important information can be estimated by computing the signal statistics, such as the covariance matrix, which requires to smooth the raw signals. Based on the CTF model \cite{talmon2009convolutive,li2019multichannel}, in narrow-band, the microphone signal is still a convolution between the source signal and the room filter, thus convolutional network seems a natural choice for modelling and processing reverberation. 

The cross-band block is composed of two frequency-convolutional modules and one full-band linear module, designed for learning cross-band spatial information. It processes each frame independently, and is shared by all frames. 
According to the bandwidth of STFT window, the adjacent frequencies highly correlate to each other \cite{avargel_SystemIdentificationShortTime_2007}, and the frequency-convolutional layers are used to learn such correlation. 
For one signal propagation path, e.g. the direct-path, the spatial features (such as IPD) for all frequencies correlate to the time difference of arrival (TDOA). Specifically, IPD is a linear function of frequency, and the slope is the TDOA. The full-band linear module applies a linear mapping onto the frequency axis to learn such linear correlations. 
The cross-band block helps to better modelling the spatial information extracted by the narrow-band block, especially for the target direct-path signal.

\addnote[sophistication]{1}{In summary, as analyzed above, spatial information are sophisticated in the sense that: (i) the direct-path (and early reflections) and late reverberation of speech, and the spatial field of ambient noise have different characteristics, which can be formulated/modelled in narrow-band and/or full-band; (ii) these information can be exploited in many different ways for conducting speech enhancement, such as spatial filtering, temporal filtering, spatial clustering, coherence test, etc.} The proposed SpatialNet is designed for learning such sophisticated spatial information, and meanwhile it is made as concise as possible, in terms of the network architecture, model size and computational complexity.
Experiments have been conducted on multiple simulated and real datasets, performing speech separation, denoising and dereverberation either individually or jointly.  
In almost all the experiments, the proposed SpatialNet achieves the state-of-the-art performance in terms of both speech quality and ASR performance. Code and audio examples for the proposed method are available at \footnote{https://github.com/Audio-WestlakeU/NBSS}.

This work is an extension of our previously published conference paper \cite{quan_MultichannelSpeechSeparation_2022}, in which we proposed the NBC network for multichannel speech separation. 
The main contributions of this work over \cite{quan_MultichannelSpeechSeparation_2022} include:
(i) we propose the new cross-band block; (ii) the network is extended for jointly performing speech separation, denoising and dereverberation, and is evaluated with much more experiments.

\textbf{Related works about full-band and narrow-band combination}. Recently, several works have been proposed also for exploiting full-band/cross-band and sub-band/narrow-band information separately and then combining them. Our previous work of FullSubNet \cite{hao2021fullsubnet} was first proposed for single-channel speech enhancement by combining full-band and sub-band spectral information. 
Based on our multichannel narrow-band LSTM network \cite{li2019waspaa}, FT-JNF \cite{tesch_InsightsDeepNonLinear_2023} flips the first LSTM layer to the frequency axis to learn cross-band information, used for multichannel speech enhancement. The proposed SpatialNet shares a similar spirit with FT-JNF, but it replaces the LSTM networks with a more powerful Conformer narrow-band block and a convolutional-linear cross-band block. 
TFGridNet \cite{wang_TFGridNetIntegratingFull_2022} also uses cross-band and narrow-band LSTM networks, plus a cross-frame self-attention module. TFGridNet is a two-stage neural beamformer, using the same cross-band and narrow-band combination network for the two stages. Compared to TFGridNet, the proposed SpatialNet has a much simpler pipeline, performing end-to-end multichannel speech enhancement. 
DasFormer \cite{wang_DasformerDeepAlternating_2023} uses self-attention for both cross-band and narrow-band processing. The proposed SpatialNet uses a convolutional-linear cross-band block, which is more functionally and computationally efficient. Overall, the proposed SpatialNet is quite different from these existing networks. Experiments show that, SpatialNet achieves either comparable or better performance than these existing networks on all tasks. 
\section{Problem Formulation}
\label{sec_problem_formulation}

\subsection{Time Domain Formulation}
For $P$ speech sources in a noisy reverberant environment, the $M$-channel microphone signals can be formulated in the time domain as:
\begin{equation}
    {x}_m(n)=\sum_{p=1}^P {y}_{pm}(n) + {e}_m(n), \ m\in\{1,...,M\}
    \label{eq_time_domain}
\end{equation}
where $m$, $p$ and $n$ denote the indices of microphone channel, speech source and discrete time, respectively. ${y}_{pm}(n)$ is the reverberant spatial image of the $p$-th source at the $m$-th microphone, ${e}_m(n)$ is ambient noise.

The spatial image ${y}_{pm}(n)$ is the convolution of the source signal $s_p(n)$ and the room impulse responses (RIR) $a_{pm}(n)$:
\begin{equation}
    {y}_{pm}(n) = a_{pm}(n) * s_p(n),
    \label{eq_time_domain_convolution}
\end{equation}
where $*$ denotes convolution. In this work, we only consider static speakers, and thus the RIRs are time-invariant. RIR is composed of direct-path, early reflections and late reverberation, and can be divided into two parts: the desired part $a_{pm}^d(n)$ and the undesired part $a_{pm}(n)-a_{pm}^d(n)$. In this work, we conduct joint speech separation, denoising and dereverberation, namely estimating the $P$ desired speech signals 
\begin{equation}
y^d_{pm}(n)= a_{pm}^d(n) * s_p(n), \ p\in\{1,...,P\}
\label{desired_time_domain}
\end{equation}
from the multichannel recordings. In practice, we only estimate the desired speech for one reference microphone, say the $r$-th microphone. Correspondingly, the desired speech signal $y^d_{pr}(n)$ will be taken as the training target of the proposed neural network. Normally, the desired RIR part is the direct-path component \addnote[early reflection]{1}{with or without some early reflections. In this paper, we take the direct-path component as the training target of the proposed method.}

\subsection{STFT Domain Formulation}
\label{sec_stft_domain}

In the STFT domain, \eqref{eq_time_domain} can be written as:
\begin{equation}
    {X}_m(f,t)=\sum_{p=1}^P {Y}_{pm}(f,t) + {E}_m(f,t), \ m\in\{1,...,M\}
\end{equation}
where ${X}_m(f,t)$, ${Y}_{pm}(f,t)$ and ${E}_m(f,t)$ are the STFT coefficients of corresponding signals, $t\in \{1,...,T\}$ and $f\in\{0,...,F-1\}$ denote the time frame and frequency indices, respectively. Based on the W-disjoint orthogonality assumption \cite{yilmaz_blind_2004}, i.e. each T-F bin is dominated by one speaker, one effective way \cite{winter_map-based_2006,boeddecker_front-end_2018} to perform speech separation is to cluster the frames using the spatial vector (will be presented latter) estimated at each frame, as different speakers have different spatial vectors.

Applying STFT to \eqref{eq_time_domain_convolution}, the time-domain convolution model is represented with the inter-frame and inter-frequency convolution model \cite{avargel_SystemIdentificationShortTime_2007, gannot_consolidated_2017}:
\begin{equation}
    {Y}_{pm}(f,t) = \sum_{f'=0}^{F-1} {A}_{pm}(f,f',t)  * S_p(f',t)
    \label{eq_inter_frame_inter_frequency}
\end{equation}
where $S_p(f,t)$ is the STFT coefficient of $s_p(n)$, and ${A}_{pm}(f,f',t)$ is a set of band-to-band ($f'=f$) and cross-band ($f'\neq f$) filters derived from ${a}_{pm}$, convolution is applied to the time axis.
Theoretically, \eqref{eq_inter_frame_inter_frequency} is fully valid when the summation over $f'$ takes all frequencies. 
However, it's shown in \cite{avargel_SystemIdentificationShortTime_2007} that a sufficiently valid approximation of ${Y}_{pm}(f,t)$ can be obtained when $f'$ only takes neighbouring frequencies in the range of $[f-l,f+l]$ (normally $l$ = 4, determined by the bandwidth of the mainlobe of
STFT window).

Although the inter-frame and inter-frequency convolution model is accurate, it is rarely used in practice due to the large complexity of cross-band filtering. Instead, the convolutive transfer function (CTF) approximation \cite{talmon2009convolutive,li2019multichannel} can be used, by discarding the cross-band filters:
\begin{equation}
    {Y}_{pm}(f,t) \approx {A}_{pm}(f,t) * S_p(f,t).
    \label{eq_CTF}
\end{equation}
Using this model, the narrowband inverse filtering techniques have been developed for speech dereverberation in \cite{li_MultichannelOnlineDereverberation_2019} based on channel equalization, and in \cite{yoshioka_GeneralizationMultiChannelLinear_2012, nakatani_UnifiedConvolutionalBeamformer_2019} based on linear prediction.

The filtering model can be further simplified with the multiplicative transfer function (MTF) approximation (also called narrow-band approximation) \cite{avargel_MultiplicativeTransferFunction_2007}:
\begin{equation}
    {Y}_{pm}(f,t) \approx {A}_{pm}(f) S_p(f,t)
    \label{eq_narrow_band_approxi}
\end{equation}
where ${A}_{pm}(f)$ is the time-invariant acoustic transfer function (ATF) of RIR, namely it is the same for all frames of one speaker. The ATF of multiple channels can form the steering/spatial vector. 
The MTF approximation is the most widely used model for its simplicity, based on which many classic speech denoising and separation techniques have been developed, such as beamforming \cite{gannot_consolidated_2017}, spatial vector clustering \cite{winter_map-based_2006, boeddecker_front-end_2018}. However, the MTF approximation is only valid when the STFT window is sufficiently larger than RIR, so it is not suitable for high reverberation scenarios. 

The desired speech signal, i.e. Eq. (\ref{desired_time_domain}), can be well modelled with the MTF model, as the desired part of RIR (the direct-path component) is shorter than the STFT window: 
\begin{equation}
    {Y}_{pm}^d(f,t) \approx {A}_{pm}^d(f) S_p(f,t),
    \label{eq_narrow_band_desired}
\end{equation}
where ${Y}_{pm}^d(f,t)$ is the STFT coefficient of $y_{pm}^d(n)$, ${A}_{pm}^d(f)$ is the Fourier transformation of ${a}_{pm}^d(n)$. In practice, for estimating the desired speech of one reference channel, instead of using the ATF itself, we actually use the relative transfer function (RTF) between other channels and the reference channel: $\tilde{{A}}_{pm}^d(f)={A}_{pm}^d(f)/A_{pr}^d(f)$.
In the free field, for the direct-path wave propagation, the RTF is $\tilde{{A}}_{pm}^d(f)=B_{pm}e^{-j 2 \pi v_f \tau_{pm}}$ in theory, where $v_f$ is the frequency in Hz, $B_{pm}$ and $\tau_{pm}$ are the frequency-independent inter-channel level and time differences, respectively, caused by the different propagation distances from the $p$-th source to the $m$-th and the $r$-th microphones. This means the RTFs of desired speech are highly correlated across frequencies. More specifically, the magnitude of RTF is the same for all frequencies, and the phase of RTF is linearly proportional to frequency. Therefore, modelling the RTFs (of desired speech) across frequencies would be helpful for increasing the modelling accuracy. This across-frequency relation of RTFs is also widely used for resolving the frequency ambiguity problem of narrow-band methods, such as in \cite{mandel_ModelBasedExpectation_2010, boeddecker_front-end_2018}. \addnote[none-free]{1}{Note that, this across-frequency relation of RTFs is still valid to a large extent in non-free field, for example, as shown in Fig. 8 of \cite{YBTASLP21} that the IPD of binaural signals is also almost linearly proportional to frequency.} From Eq. (\ref{eq_narrow_band_desired}), it is obvious that the desired speech is spatially coherent. 

The undesired signals to be removed mainly consists of the late reverberation and ambient noise. Ambient noise is random signal with either time-invariant (stationary) or time-variant (non-stationary) power spectrum. 
Spatially, the sound field for both late reverberation and ambient noise are (partially) diffuse \cite{gannot_consolidated_2017,habets2008generating}, with the spatial correlation between two microphones of  
\begin{equation}
    {r}(f,d) = \frac{sin(2\pi v_f d/c)}{2\pi v_f d/c},
    \label{eq_diffuse_correlation}
\end{equation}
where $d$ is the distance of the two microphones, $c$ denotes the propagation speed of sound.  The spatial correlation is a \emph{sinc} function of $f$ and $d$, and it gradually decreases with the increase of $f$ and $d$. In \cite{mohan2008localization}, a coherence test method is proposed for detecting direct-path dominated frames based on the difference of spatial correlation for the desired (direct-path) signal and the undesired reverberation and noise.

Overall, the above mentioned methods, i.e. narrowband inverse filtering, beamforming, T-F bin clustering and coherence test, all largely leverage the narrow-band spatial information to conduct speech enhancement. Narrowband indeed involves rich information for discriminating one desired speech from other signals, such as from other desired speech with their different RTFs (steering vectors), from reverberation either by identifying the CTFs (narrowband room filters) or with their different spatial correlations, from noise with their different spatial correlations. Besides the narrow-band information, leveraging the cross-frequency information would also be helpful for increasing the modelling accuracy of desired speech. Based on these analysis, we will propose our SpatialNet in the next section.

\section{SpatialNet}
\label{sec_spatialnet}

\begin{figure*}[htbp]
  \centering
  \includegraphics[width=0.95\linewidth]{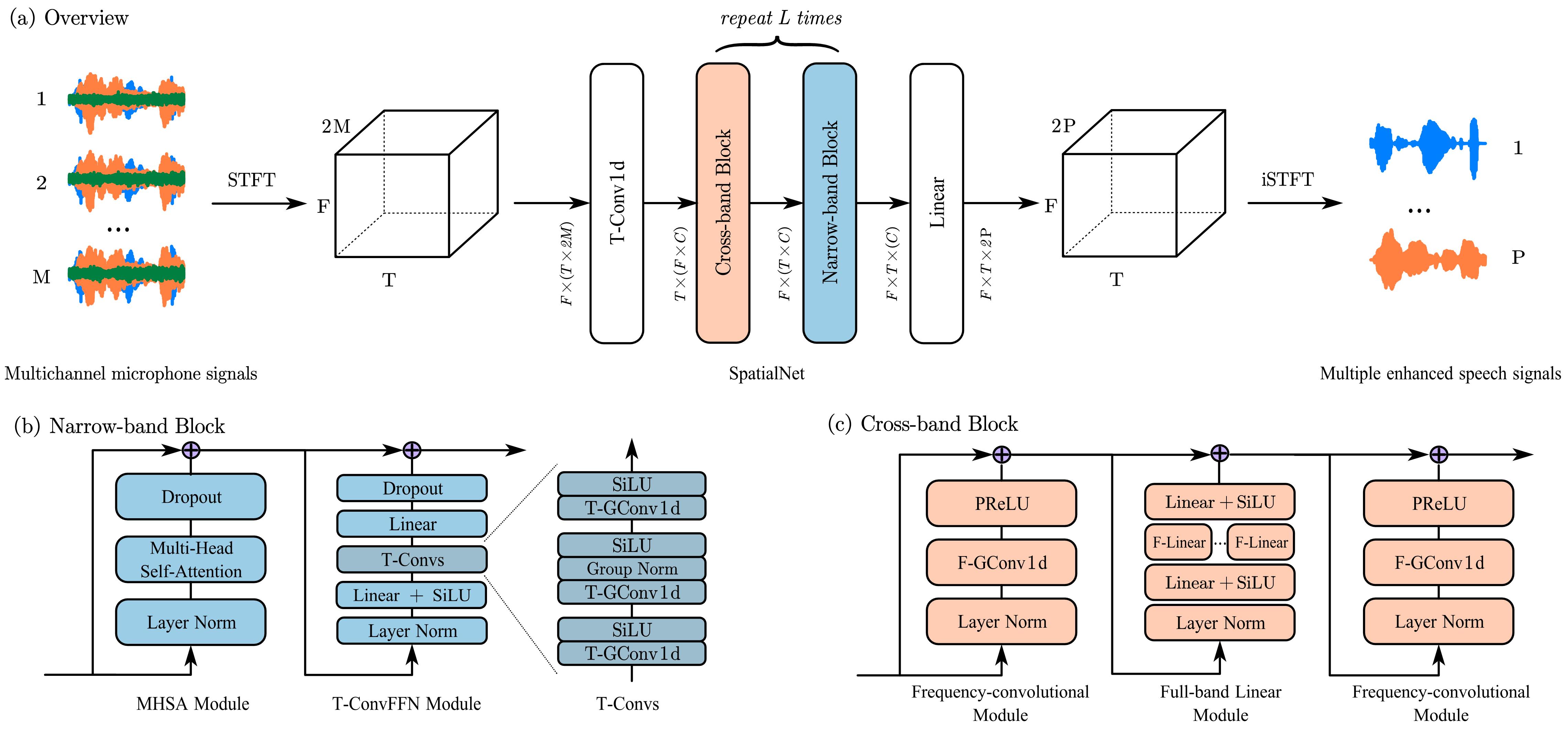}
  \caption{The proposed SpatialNet. (a) The system overview. The input dimensions of neural blocks are presented before each of them in the form ``batch dimension $\times$ (dimension of one sample in batch)".  (b) The narrow-band block. (c) The cross-band block.}
  \label{fig_arch_spatial_net}
 \vspace{-0.2cm}
\end{figure*}

This section presents the proposed SpatialNet. It is designed for learning the sophisticated spatial information mentioned in the previous section, and a lot of efforts have been made to maintain the network architecture as simple as possible.
Fig. \ref{fig_arch_spatial_net} (a) shows the systematic overview of \addnote[spatialnet]{1}{SpatialNet}.
Before/after the network processing, the time domain waveforms are transformed to/from STFT coefficients.
The input of the network is formed by concatenating the real and imaginary parts of multichannel microphone signals for each T-F bin:
\begin{equation}
\begin{aligned}
    {\mathbf{x}}[f,t,:] =  [ \mathcal{R}&(X_1(f,t)), \mathcal{I}(X_1(f,t)), ...,\\ & \mathcal{R}(X_M(f,t)), \mathcal{I}(X_M(f,t)) ] \in \mathbb{R}^{2M} \notag
\end{aligned}
\end{equation}
where $[:]$ is an operator to take all values of one dimension of a tensor, and $\mathcal{R}(\cdot)$ and $\mathcal{I}(\cdot)$ denote the real and imaginary parts of complex number, respectively. 
The network output is the prediction of concatenated STFT coefficients of all desired speech signals for each T-F bin:
\begin{equation}
\begin{aligned}
    {\mathbf{y}}[f,t,:] = [ \mathcal{R}&(Y_{1r}^d(f,t)), \mathcal{I}(Y_{1r}^d(f,t))), ...,\\ &\mathcal{R}(Y_{Pr}^d(f,t))), \mathcal{I}(Y_{Pr}^d(f,t))) ] \in \mathbb{R}^{2P} \notag
\end{aligned}
\end{equation}

As shown in Fig. \ref{fig_arch_spatial_net} (a), SpatialNet is composed of one convolutional input layer (T-Conv1d), $L$ interleaved narrow-band blocks (see Section \ref{sec_narrow_band}) and cross-band blocks (see Section \ref{sec_cross_band}), and one linear output layer.
The convolutional input layer conducts convolution on $\mathbf{x}$ with a kernel size of 1 and 5 on the frequency and time axes, respectively, and with $C$ channels, obtaining a hidden representation $\mathbf{h}_0\in \mathbb{R}^{F\times T\times C}$. Then, it will be processed by the interleaved cross-band and narrow-band blocks.
The Linear output layer maps the output of the last block to the concatenated target STFT coefficients ${\mathbf{y}}$. 
Finally, the time-domain speech signals can be obtained by applying inverse STFT. 
For network training, the loss function is set as the negative of the scale-invariant signal-to-distortion ratio (SI-SDR) \cite{roux_sdr_2019} of the time-domain enhanced speech signals. 
Permutation invariant training is adopted for solving the label permutation problem  \cite{yu_permutation_2017}.

Along the entire network, the input and output of one layer/block will always be a hidden tensor with the dimension of $F\times T\times C$. For notational simplicity, hereafter, we omit the layer/block index for the hidden tensors, and denote them as $\textbf{h}\in \mathbb{R}^{F\times T\times C}$ whenever there is no ambiguity.

\subsection{Narrow-band Block}
\label{sec_narrow_band}

As presented in Section \ref{sec_stft_domain}, narrow-band involves rich information for speech enhancement, and the narrow-band block is designed for learning such information. The narrow-band block processes each frequency independently, and all frequencies share the same network parameters. 
At one frequency, the frames dominated by different speakers (and non-speech signals) can be clustered based on their spatial vectors \cite{winter_map-based_2006,boeddecker_front-end_2018}.
From the perspective of computing the similarity of vectors, spatial vector clustering shares a similar principle with the self-attention mechanism, which motivates us to employ self-attention in our narrow-band block. \addnote[attn spk track]{1}{In a similar spirit, the self-attention mechanism is also used in \cite{ochiai_MaskBasedNeuralBeamforming_2023} for measuring the similarity of the instantaneous spatial covariance matrix of different time frames.}

Speech and noise signals are random processes, and the estimation of spatial features relies on the computation of signal statistics, such as the covariance matrix. This motivates us to use convolution layers to perform local smoothing/averaging operations for the computation of signal statistics. In addition, based on the CTF model shown in Eq.~\eqref{eq_CTF}, in narrow-band, the microphone signal is still a convolution between the source signal and CTF. Thus, it seems a natural choice to use convolutional layers along the time axis for modelling and processing reverberation. For example, the convolutional layers may imitate the inverse filtering process in the way of linear prediction \cite{yoshioka_GeneralizationMultiChannelLinear_2012, nakatani_UnifiedConvolutionalBeamformer_2019}.

As shown in Fig. \ref{fig_arch_spatial_net} (b), the narrow-band block is composed of one multi-head self-attention (MHSA) module and one time-convolutional feed forward network (T-ConvFFN). This block works on a single STFT frequency, and the frequency axis is taken within the batch dimension. 

\subsubsection{Multi-head Self-attention Module}
\label{sec:MHSA}
This module is designed to collect/separate the components of the same/different speakers using the self-attention technique \cite{vaswani_attention_2017} by computing the similarity of spatial vectors in one STFT frequency.
It consists of a Layer Norm (LN) \cite{ba_LayerNormalization_2016}, a standard Multi-Head Self-Attention (MHSA) \cite{vaswani_attention_2017}, a dropout, and a residual connection from the module input to the module output.
This module is formulated as:
\begin{equation}
    \textbf{h}[f,:,:] \leftarrow \textbf{h}[f,:,:] +  \text{Dropout(MHSA(LN}(\textbf{h}[f,:,:]))) \notag
\end{equation}
for all $f$'s.

\subsubsection{T-ConvFFN Module}
\label{sec:ConvFFN}
This module modifies the feed forward network used in Transformer \cite{vaswani_attention_2017} by inserting time-convolutional layers (denoted as T-Convs) in between the two linear layers.
The whole module is formulated as:
\begin{equation}
\begin{aligned}    
    \textbf{h}[f,:,:] \leftarrow & \textbf{h}[f,:,:] + \text{Dropout(Linear(T-Convs(}\\ &\text{SiLU(Linear(LN}(\textbf{h}[f,:,:])))))). \notag
\end{aligned}
\end{equation}
for all $f$'s, where the first linear layer transforms the hidden vector from $C$-dim to a higher dimension, say $C'$, while the last linear layer transforms the hidden vector from $C'$-dim back to $C$-dim.
The Sigmoid Linear Unit (SiLU) activation function  \cite{hendrycks_gaussian_2020, ramachandran_searching_2017} is applied after the first linear layer and after the convolutional layers in T-Convs.

In T-Convs, three group convolutional layers are used and group normalization (GN) \cite{wu_group_2018} is applied after the second convolutional layer.
The group convolutional layers perform 1-D convolution along the time dimension.
The number of channels for the convolutional layers is $C'$, and the channels are split into $G$ groups. 

Compared to the convolutional network used in Conformer proposed in \cite{gulati2020conformer}, the major differences of the proposed T-ConvFFN include (i) \cite{gulati2020conformer} uses one convolutional layer, and we use three convolutional layers; \addnote[largercn]{1}{(ii) \cite{gulati2020conformer} puts the convolutional layer before the feed-forward module, and the number of channels is $C$, while we put the convolutional layers in between the two feed-forward linear layers, thus time convolution is conducted on a larger number of channels, i.e. $C'$.} These differences account for the high requirement of local smoothing/averaging and reverberation processing for narrowband signal modelling.

\subsection{Cross-band Block}
\label{sec_cross_band}
As introduced in Section \ref{sec_stft_domain}, due to the STFT window effect, the microphone signal of ${Y}_{pm}(f,t)$ for frequency $f$ contains the information of the source signal $S_p(f,t)$ for $l$ (about 4) neighbouring frequencies \cite{avargel_SystemIdentificationShortTime_2007}. 
Reversely, the neighbouring frequencies of microphone signals would also be helpful for the estimation of source signal.
Besides, the spatial feature of desired speech, i.e. the RTF of direct-path speech, is highly correlated across all frequencies. 
Accordingly, we propose the cross-band block for learning these cross-band spatial information, including two frequency-convolutional layers and one full-band linear module, which are shown in Figure \ref{fig_arch_spatial_net} (c). This cross-band block processes each time frame independently, and all time frames use the same network. In other words, this block works on a single STFT frame, and the time frame axis is taken within the batch dimension.

\subsubsection{Frequency-convolutional Module}
The frequency-convolutional module is proposed to model the correlation between neighbouring frequencies.
This module consists of one LN, one group convolution along frequency axis (F-GConv1d) and one parametric ReLU (PReLU) activation unit \cite{he_DelvingDeepRectifiers_2015}. It can be formulated as:
\begin{equation}
    \textbf{h}[:,t,:] \leftarrow \textbf{h}[:,t,:] + \text{ PReLU(F\mbox{-}GConv1d(LN}(\textbf{h}[:,t,:]))) \notag
\end{equation}
for all $t$'s.

\subsubsection{Full-band Linear Module}
Due to the existence of interference signals, e.g. interfering speakers, reverberation and noise, it is difficult to accurately model the spatial feature of desired speech in narrow-band. Leveraging the spatial feature correlation across frequencies would be helpful, which motivates us to propose the full-band module.
In the full-band linear module, we first use a Linear layer with SiLU activation (shared by all T-F bins) to reduce the number of hidden channels to $C''$:
\begin{equation}
    \textbf{h}'[f,t,:] = \text{SiLU(Linear(}\textbf{h}[f,t,:])) \in \mathbb{R}^{C''} \notag
\end{equation}
for all $f$'s and $t$'s.
Then, full-band mapping is conducted onto the frequency axis by a group of linear networks. Different hidden channels use different linear networks, denoted as $\text{F-Linear}_c$ for $c=1,...,C''$. It is formulated as
\begin{equation}
    \textbf{h}'[:,t,c] \leftarrow \text{F-Linear}_c(\textbf{h}'[:,t,c]) \notag
\end{equation}
for all $t$'s and $c$'s.
Finally, the output of the module is obtained by increasing the number of channels to $C$  using a Linear layer with SiLU activation, then add the original input of this module:
\begin{equation}
    \textbf{h}[f,t,:] \leftarrow \textbf{h}[f,t,:] +  \text{SiLU(Linear(}\textbf{h}'[f,t,:]))
\end{equation}
for all $f$'s and $t$'s.

In this module, $C''$ is several times smaller than $C$, and $F$ is comparable to or larger than $C$. Thus, the parameters and computations of this module mainly lie in the F-Linear networks.
To be parameter efficient, the same F-Linear networks are shared for all the repeated cross-band blocks. It is interesting to find from our preliminary experiments that sharing the F-Linear networks for all cross-band blocks does not degrade the performance.

\subsection{Discussions}
\label{sec:discussions}
The proposed SpatialNet is designed for learning narrow-band and cross-band spatial information. However, it is difficult to fully disentangle spatial information from spectral information. The narrow-band block processes frequencies independently, but one frequency still involves some spectral information. For example, the narrow-band spectral evolution/envelope of speech is quite different from the one of noise, as speech is more temporally correlated and non-stationary, while noise is \addnote[typi]{1}{typically} less temporally correlated and stationary. The narrow-band block may also learn this spectral difference for denoising. The full-band mapping networks see all frequencies, thus may learn the full-band spectral pattern of signals, but we believe that it does not learn much, as the cross-band block processes frames independently, and one frame does not include any spectral dynamics. In addition, the representation capabilities of the full-band mapping networks are limited, as they are shared for all cross-band blocks. This issue will be verified in Section~\ref{sec:SpectralInformation}.

Regarding the order of narrow-band and cross-band blocks, our designing order is first narrow-band then cross-band. As demonstrated by many narrow-band techniques, e.g. beamforming, narrow-band inverse filtering and frame clustering, narrow-band provides fundamental information, while cross-band provides some auxiliary information. Our experiments (Section~\ref{sec:sub-network}) also testify that, in our designed SpatialNet, compared to the cross-band block, the narrow-band block contributes much more to the performance, and also consumes much more computations. The cross-band block is put before the narrow-band blocks, since this order achieves somewhat better performance than the other way around. However, the two blocks will be repeated many times, so the order of them does not matter much.

\section{Experiments}

\subsection{Experimental Setups}

\subsubsection{Network Configuration} For the proposed network, we set the kernel size of the convolutional input layer (T-Conv1d), time-dimension group convolution (T-GConv1d), and frequency-dimension group convolution (F-GConv1d) to 5, 5 and 3, respectively.
The group numbers of T-GConv1d, F-GConv1d and GN, i.e. $G$, are all set to 8.
The number of self-attention heads is set to 4.
A small version and a large version of the proposed network are proposed/suggested. 
The small network, referred to as \textbf{SpatialNet-small}, uses $L=8$ blocks; and the numbers of hidden units are set to $C=96$, $C'=192$, and $C''=8$.
The large network, referred to as \textbf{SpatialNet-large}, uses $L=12$ blocks; and the numbers of hidden units are set to $C=192$, $C'=384$ and $C''=16$.

STFT is applied using Hanning window with a length of 512/256 samples (32ms) and a hop size of 256/128 samples (16ms) for the 16/8 kHz data.
The number of frequency bins processed by the network is 257/129 for the 16/8 kHz data. The model size is dependent on the number of frequencies due to the full-band mapping networks. Specifically, for the 16/8 kHz data, the model size of SpatialNet-small is 1.6/1.2 M, and the model size of SpatialNet-large is 7.3/6.5 M. 

As for network training, the batch size is set to 2 utterances. \addnote[batch size]{1}{In our preliminary experiments, 2, 4 and 8 utterances have all been tried (the learning rate for the three cases are the same one as will be described later), and their performance are very close. The narrow-band/cross-band blocks in the proposed network process frequencies/frames independently, and are shared by all frequencies/frames. Thence, different frequencies/frames can be somewhat considered as independent training samples for the narrow-band/cross-band blocks, and a few utterances may provide sufficient training samples for one batch.} The length of training utterances are always 4 seconds. The Adam \cite{kingma2015adam} optimizer is used with a learning rate initialized to 0.001 and exponentially decayed as $lr \xleftarrow{} 0.001 * 0.99^{epoch}$. Gradient clipping is applied with a gradient norm threshold of 5.

\subsubsection{Datasets}
The proposed network is developed for joint speech separation, denoising and dereverberation.
However, there are few public datasets designed for the joint task.
We evaluate the proposed network with six widely used public datasets conducting either one or two of the three tasks, or the joint three tasks, including SMS-WSJ \cite{drude_SMSWSJDatabasePerformance_2019}, WHAMR! \cite{maciejewski_WHAMRNoisyReverberant_2020}, Spatialized WSJ0-2mix \cite{wang_multi-channel_2018}, LibriCSS \cite{chen_ContinuousSpeechSeparation_2020}, Reverb Challenge \cite{kinoshita_ReverbChallengeCommon_2013} and CHiME 3/4 Challege \cite{barker_ThirdCHiMESpeech_2015}. The former three are simulated two-speaker mixture datasets, while the latter three are real-recorded datasets.  

The proposed SpatialNet performs end-to-end multichannel speech enhancement in the STFT domain, thus it is microphone-array-dependent. For different datasets, the networks are independently trained using the training data recorded/simulated with their specific microphone arrays. \addnote[ref-mic]{1}{If not otherwise stated, the first channel (according to the microphone order of each dataset) is chosen as the reference microphone for both training and evaluation.}

\subsubsection{Evaluation Metrics} 
The speech enhancement performance is evaluated in terms of both speech quality and ASR. As for speech quality, we use the widely used metrics, including signal-to-distortion ratio (SDR) \cite{vincent_performance_2006}, scale-invariant SDR (SI-SDR) \cite{le2019sdr},
narrow-band or wide-band perceptual evaluation of speech quality (NB- or WB-PESQ) \cite{rix_perceptual_2001}, short-time objective
intelligibility (STOI) \cite{taal2010short} and extended STOI (eSTOI) \cite{jensen2016algorithm}. For all these metrics, the larger the better. As for ASR,  Word Error Rate (WER) is used as the evaluation metric, for which the smaller the better.
One exception is that, for the Reverb Challenge dataset, we use the official evaluation metrics of the challenge. 

\subsubsection{Comparison Methods}
As experiments are conducted on multiple datasets and tasks, it is not easy to reproduce other methods with proper configurations. Therefore, for each dataset, we compare with the methods that have reported their results on the dataset, and if not otherwise stated their results are directly quoted from their original paper. We have carefully searched the literature to involve as much as possible recently proposed SOTA methods for comparison.  

\subsection{Ablation Studies}
\label{sec_ablation}

To analyze the characteristics of the proposed network, we conduct ablation experiments on an extended SMS-WSJ dataset. SMS-WSJ \cite{drude_SMSWSJDatabasePerformance_2019} is a simulated two-speaker mixture dataset.
Clean speech signals are sampled from the Wall Street Journal (WSJ) corpus.
A six-microphone circular array with a radius of 10 cm is simulated.
RIRs are generated using the image method \cite{allen_image_1979}.
The reverberation time (T60) is uniformly sampled from 0.2 s to 0.5 s.
The source positions are randomly sampled being [1, 2] m away from the microphone array center.
Artificially generated white sensor noise is added to speech mixtures with a signal-to-noise ratio (SNR) uniformly sampled in the range of [20, 30] dB.
The sampling rate is 8 kHz.
A baseline ASR model is provided based on Kaldi \cite{Povey_Kaldi_ASRU2011}.

The original SMS-WSJ dataset is extended by introducing larger reverberation and noise to evaluate the proposed network in more adverse environments. The extended dataset is named SMS-WSJ-Plus. All the six microphones are used. 
T60 is extended to [0.1, 1.0] s.
The speaker-to-microphone distance is extended to [1, 4] m.
Two speech signals are mixed together with a signal-to-interference ratio (SIR) randomly sampled in [-5, 5] dB.
Multichannel diffuse babble or white noise generated using \cite{habets2008generating} is added to the speech mixture with a SNR sampled in [0, 20] dB. 

\subsubsection{Contribution of Sub-networks}
\label{sec:sub-network}

Table~\ref{table_sms_wsj_plus} shows the enhancement performance, the number of parameters and the number of floating point operations (FLOPs) \footnote{FLOPs in Giga per second (G/s) is measured with four-second long utterance and then divided by four, as we normally process four-second long signals in this work. We use the official tool provided by PyTorch (torch.utils.flop\_counter.FlopCounterMode on meta device) for FLOPs computation.\label{flops}} of SpatialNet-small and its variants by removing or replacing each sub-network of it. It can be seen that every sub-networks noticeably contribute to the enhancement performance. The narrow-band modules, i.e. MHSA and T-ConvFFN, have 0.3 M and 0.6 M parameters, and 11.0 G/s and 10.1 G/s FLOPs, respectively. By contrast, the two cross-band modules, i.e. frequency-convolutional modules and full-band linear module, have 0.1 M and 0.15 M parameters, and 1.5 G/s and 0.3 G/s FLOPs, respectively. Correspondingly, \addnote[design]{1}{in our designed SpatialNet, the narrow-band block plays a more fundamental role, and contributes more than the cross-band block.}

\begin{table}[tpb]
\setlength\tabcolsep{2pt}
\renewcommand{\arraystretch}{1.3}
\caption{Contribution of sub-networks, experiments on SMS-WSJ-Plus.}
\label{table_sms_wsj_plus}
\centering
\resizebox{\linewidth}{!}{
\begin{tabular}{lcccccc}
\hline\hline
Network & \#Param & FLOPs & NB-PESQ & ESTOI & SI-SDR   & WER  \\
Architecture &  (M) & (G/s) &       &       &  (dB)    &  (\%)  \\
\hline
unproc. & - & - & 1.35 & 0.261 & -11.4 & 91.13 \\ %
\hline
SpatialNet-Small & 1.2  & 23.1  & 3.11 & 0.878 & 13.8 & 15.89 \\ %
\ remove MHSA  & 0.9  & 12.1  & 2.55 & 0.777 & 9.5 & 31.62 \\ %
\makecell[l]{\ replace T-ConvFFN with \\\ \ \ Transformer FFN} & 0.9  & 17.7  & 2.67 & 0.796 & 10.5 & 26.91 \\ %
\ remove Frequency-convolutional  & 1.1  & 21.6  & 3.03 & 0.863 & 12.8 & 17.01 \\ %
\ remove Full-band Linear Module & 1.0  & 22.8  & 2.80 & 0.838 & 12.1 & 20.99 \\ %
\hline\hline
\end{tabular}
}
\vspace{-0.2cm}
\end{table}

\subsubsection{Attention Map Analysis}

As shown in Table~\ref{table_sms_wsj_plus}, the narrow-band MHSA module contributes the most to the speech enhancement performance. It is expected to perform spatial clustering of different speakers and of non-speech frames. To verify this, in Fig. \ref{fig:attn}, we draw the attention maps of one example utterance. The first three rows and the first three columns of Fig. \ref{fig:attn} are the spectrogram of clean speech signal of the first speaker (`spk1'), the second speaker (`spk2') and their noisy reverberant mixture (`mix'), respectively. 
Let's denote the attention score of one head for all frequencies as $\text{Attention}_{f,q,k} \in [0,1]$, where $f\in\{0,...,F-1\}$, $q\in\{1,...,T\}$ and $k\in\{1,...,T\}$ denote the indices of frequency, query and key, respectively, with $\sum_k \text{Attention}_{f,q,k}=1$ according to the softmax function along $k$. 
In the fourth row, we draw the Q-K (Query-Key) attention maps of two representative heads in the fourth and fifth columns, respectively. The Q-K attention maps draw the attention scores averaged over frequencies, i.e. $\frac{1}{F}\sum_f \text{Attention}_{f,q,k}$, and reflect the attentions between frames. 
In the fifth row, we draw the F-K (Frequency-Key) attention maps of the same two heads. The F-K attention maps draw the attention scores averaged over queries, i.e. $ \frac{1}{T}\sum_q \text{Attention}_{f,q,k}$, and reflect the contribution of each T-F bin (to other T-F bins at the same frequency).
We can see that the two heads model the non-speech signals and (direct-path) speech signals, respectively. In the first head (the fourth column), the non-speech T-F bins are attended (shown by the F-K map) and the attentions are temporally global (shown as the vertical bars in the Q-K map). In the second head (the fifth column), only the speech T-F bins are attended, as shown by the F-K map. Moreover, two speakers are well clustered as shown in the Q-K map that the speech frames of one speaker only attend to the speech frames of the same speaker. This verifies that the narrow-band MHSA module is indeed performing speaker (and non-speech) clustering based on narrow-band spatial features. 

\begin{figure}[tbp]
  \centering
  \includegraphics[width=1\linewidth]{uttr_0_T60_0.40_SNR_10.00_SIR_0.78_babble.jpeg}
  \caption{Attention maps of one example utterance.}
  \label{fig:attn}
  \vspace{-0.4cm}
\end{figure}

\subsubsection{Exploitation of Spectral Information } 
\label{sec:SpectralInformation}

The proposed SpatialNet is proposed to extensively leverage the spatial information.
However, as mentioned in Section~\ref{sec:discussions}, it may also learn some spectral information.
In order to investigate the extent to which spectral information is utilized, we conduct three ablation experiments on the six-channel SMS-WSJ-Plus dataset. An advanced two-stage neural beamformer, i.e. Beam-Guided TasNet \cite{chen_BeamGuidedTasNetIterative_2022}, is compared\addnote[official beam]{1}{ using it's official implementation\footref{bg code}}. It interleaves an end-to-end speech enhancement neural network (i.e. MC-TasNet \cite{gu_EndtoEndMultiChannelSpeech_2019}) and the MVDR beamformer. The beamformer is spectral-agnostic, while the end-to-end network learns both spectral and spatial information. It may not be very suitable for Beam-Guided TasNet to process the highly-reverberant SMS-WSJ-Plus dataset, as the MVDR beamformer is designed not for dereverberation, and does not work well for high-reverberation scenarios. So, Beam-Guided TasNet mainly serves here as a baseline method for evaluating the exploitation of spectral information by different methods.

\addnote[information]{1}{\textbf{Same Speaker Position.}
In the first experiment, we put two speakers at the same position (using the same RIR for the two speakers), such that the network can only rely on spectral information to separate the two speakers.  Note that, the two speakers are always put at different positions in training, and can be put at the same position for test. Table \ref{table_spatial_for_separation} shows the results. 
For the six-channel case, when the two speakers are put at the same position, the performance of both Beam-Guided TasNet and the proposed network degrade, but the performance degradation of the proposed network is much more significant, which means the proposed network is much more relying on spatial information. By listening to the enhanced signals, when the two speakers are put at the same position, noise and reverberation can still be largely suppressed (thus the performance measures are still improved relative to the unprocessed signals), but the two speakers almost cannot be separated, which indicates that the proposed SpatialNet leverages little spectral information for speech separation.
}

\addnote[single-channel]{1}{\textbf{Single-channel Speech Enhancement.} We also train the proposed SpatialNet for single-channel speech enhancement, to test how the network behaves when less spatial information can be used. The two speakers are also put at different positions in training, and can be put at the same position for test. The results are also shown in Table \ref{table_spatial_for_separation}. It is interesting to find that the proposed SpatialNet also performs well for single-channel speech enhancement, although it is designed for multi-channel speech enhancement. When the two speakers are put at the same position, the performance measures also significantly degrade for the single-channel case, which indicates that speech separation is conducted by exploiting the single-channel spatial difference (RIR difference) of the two speakers. This is somehow out of expectation as the network needs to blindly identify the RIR information for each speaker from single-channel speech signal, which is very difficult and has rarely been studied in the field as far as we know. When two speakers are put at different positions, the self-attention maps for single-channel enhancement exhibit similar speaker clustering characteristics as for the multi-channel case shown in Fig. \ref{fig:attn}, which further verifies that the proposed SpatialNet can perform single-channel speech separation based on spatial clustering. Compared to the six-channel case, the single-channel case leverages more spectral information, as its performance degradation is smaller when the two speakers are put at the same position. 
}      

\begin{table}[tbp]
\setlength\tabcolsep{3pt}
\renewcommand{\arraystretch}{1.2}
\caption{Experiments of putting two speakers at the same position and single-channel speech enhancement.}
\label{table_spatial_for_separation}
\centering
\resizebox{1\linewidth}{!}{
\begin{tabular}{cccccc}
\hline\hline
\multirow{1}{*}{\textbf{Network}} & \multirow{1}{*}{\textbf{\#CH}} & \textbf{Same Position} & NB-PESQ & ESTOI & SDR(dB) \\
 \multirow{2}{*}{unproc.} & \multirow{2}{*}{-}
 & \xmark & 1.37 & 0.262 & -2.7 \\
 & & \cmark & 1.37 & 0.261 & -2.7 \\
\hline
\hline
\multirow{2}{*}{\makecell[c]{Beam-Guided TasNet\\\ (iter=2) \cite{chen_BeamGuidedTasNetIterative_2022}}} & \multirow{2}{*}{6}
 & \xmark & 1.98 & 0.606 & 8.0 \\
 & & \cmark & 1.65 & 0.481 & 4.2 \\
\hline
\multirow{2}{*}{\makecell[c]{SpatialNet-small (prop.)}} & \multirow{2}{*}{6}
 & \xmark & 3.04 & 0.857 & 15.1 \\
 & & \cmark & 1.69 & 0.552 & 1.8 \\
\hline
\multirow{2}{*}{\makecell[c]{SpatialNet-small (prop.)}} & \multirow{2}{*}{1}
 & \xmark & 2.27 & 0.690 & 9.2\\ 
 & & \cmark & 1.73 & 0.541 & 3.3 \\ 
\hline
\hline
\end{tabular}
}
\vspace{-0.2cm}
\end{table}

\textbf{Spectral Generalization Experiment.}
If the network learns spectral information, it would suffer from the spectral generalization problem. 
In this experiment, the spectral generalization ability of the proposed SpatialNet is evaluated by performing cross-language speech enhancement, as the spectral pattern of different languages are quite different. 
SMS-WSJ-Plus is used here as an English dataset.
A Chinese dataset is constructed by simply replacing the clean speech signals of SMS-WSJ-Plus with (our private) clean Chinese speech signals. 
The length of test utterances in the two datasets are all set to four seconds.
Table \ref{table_lang_gen} reports the results.
We can see that both the proposed SpatialNet and Beam-Guided TasNet have certain cross-language generalization problem, as the performance measures degrade when training with different language as the test data. It is obvious that the performance degradation of the proposed SpatialNet is smaller than the one of Beam-Guided TasNet.
As an anchor, we can analyze the `English' test case, for which SpatialNet achieves similar ESTOI and SDR scores when trained with `English' or `Chinese' data, while Beam-Guided TasNet achieves much lower scores trained with `Chinese' than `English' data. 
Overall, the performance degradation of the proposed SpatialNet is mild for cross-language speech enhancement, which indicates that the network dominantly leverages spatial information over spectral information.

\begin{table}[tbp]
\setlength\tabcolsep{3pt}
\renewcommand{\arraystretch}{1.3}
\caption{Cross-language speech enhancement results. NB-PESQ, ESTOI and SDR (dB) are reported in the form of "NB-PESQ/ESTOI/SDR".}
\label{table_lang_gen}
\centering
\resizebox{1\linewidth}{!}{
\begin{tabular}{ccccc}
\hline\hline
Test Language & \multicolumn{2}{c}{English} & \multicolumn{2}{c}{Chinese} \\
Training Language & English & Chinese & English & Chinese \\
\hline
unproc. & \multicolumn{2}{c}{1.37/0.262/-2.7} & \multicolumn{2}{c}{1.20/0.302/-2.8} \\
\hline
\makecell[l]{Beam-Guided TasNet\\\ (iter=2) \cite{chen_BeamGuidedTasNetIterative_2022}}
  & 1.98/0.606/8.0 & 1.76/0.485/5.5 & 1.53/0.507/4.8 & 1.68/0.604/6.9 \\
SpatialNet-small (prop.) & 3.04/0.857/15.1 & 2.87/0.839/14.6 & 2.76/0.877/15.9 & 2.94/0.875/15.9 \\
\hline\hline
\end{tabular}
}
\vspace{-0.2cm}
\end{table}

\subsection{Results on SMS-WSJ}

We evaluate the proposed networks on the original SMS-WSJ dataset, and compare with other methods. 
Two-channel and six-channel results are reported in Table \ref{table_smswsj}. For the comparison methods, if the results of several variants have been reported in their papers, we quote the results of the best variant.  
From the table, we can see that both speech quality and ASR performance can be largely improved by the speech enhancement methods. The time-domain end-to-end networks, i.e. FaSNet$+$TAC \cite{luo_end--end_2020} and Multi-TasNet \cite{zhang_mc_convtasnet_2020}, don't perform as well as other methods. Other comparison methods, i.e. MISO$_1$-BF-MISO$_3$ \cite{wang_MultimicrophoneComplexSpectral_2021}, Convolutional Prediction \cite{wang_ConvolutivePredictionReverberant_2021}, MC-CSM with LBT \cite{taherian_LBT_2022} and TFGridNet \cite{wang_TFGridNetIntegratingFull_2022}, all perform neural beamforming plus neural post-processing, and achieve much better ASR performance than the time-domain end-to-end networks. This demonstrates the advantage of combining beamforming and deep learning techniques. It is also consistent to the widely agreed view in the field that beamforming (linear spatial filtering) is more friendly to ASR, compared to end-to-end neural speech enhancement. Among the comparison methods, TFGridNet performs the best, by adopting an advanced full-band and sub-band combination network. 

Compared to TFGridNet, the proposed SpatialNet-large achieves better speech enhancement performance and comparable ASR performance. This demonstrates that, by extensively exploit the narrow-band and cross-band spatial information, target direct-path signals can be well recovered from very noisy microphone recordings. In addition, our (STFT-domain) end-to-end speech enhancement network is efficient for improving both speech quality and ASR performance. SpatialNet-small also achieves very good performance with much less parameters and computations. 

\begin{table}[tpb]
\setlength\tabcolsep{3pt}
\renewcommand{\arraystretch}{1.3}
\caption{Results on 2-channel and 6-channel SMS-WSJ Dataset. 
$^{\star}$ 
denote that the scores are quoted from \cite{wang_MultimicrophoneComplexSpectral_2021}.
}
\label{table_smswsj}
\centering
\resizebox{\linewidth}{!}{
\begin{tabular}{lccccc}
\hline\hline
\textbf{Method} & SISDR (dB) & SDR (dB) & NB-PESQ & eSTOI & WER (\%)\\
\hline
unproc. & -5.45 & -0.38 & 1.50 & 0.441 & 78.7 \\ 
\hline
oracle direct-path  & $\infty$ & $\infty$ & 4.5 & 1.0 & 6.31 \\
\hline\hline
\multicolumn{6}{c}{\textbf{Results for 2-channel SMS-WSJ}}\\
\hline
FasNet+TAC $^{\star}$ \cite{luo_end--end_2020} & 6.9 & - & 2.27 & 0.731 & 34.84 \\
Multi-TasNet $^{\star}$ \cite{zhang_mc_convtasnet_2020} & 5.8 & - & 2.16 & 0.720 & 45.72 \\
MISO$_1$-BF-MISO$_3$ \cite{wang_MultimicrophoneComplexSpectral_2021} & 12.7 & - & 3.43 & 0.907 & 10.67 \\
Convolutional Prediction \cite{wang_ConvolutivePredictionReverberant_2021} & 15.8 & - & 3.71 & - & 8.60 \\
TFGridNet \cite{wang_TFGridNetIntegratingFull_2022} & 20.3 & 22.0 & 3.81 & 0.967 & 7.41 \\
\hline
SpatialNet-small (prop.)  & 19.4 & 21.0 & 3.80 & 0.957 & 8.10 \\
SpatialNet-large (prop.)  & \textbf{23.3} & \textbf{24.6} & \textbf{4.03} & \textbf{0.975} & \textbf{7.20} \\
\hline\hline
\multicolumn{6}{c}{\textbf{Results for 6-channel SMS-WSJ}}\\
\hline
FasNet+TAC $^{\star}$ \cite{luo_end--end_2020} & 8.60 & - & 2.37 & 0.771 & 29.8 \\
Multi-TasNet $^{\star}$ \cite{zhang_mc_convtasnet_2020} & 10.8 & - & 2.78 & 0.844 & 23.1 \\
MISO$_1$-BF-MISO$_3$ \cite{wang_MultimicrophoneComplexSpectral_2021} & 15.6 & - & 3.76 & 0.942 & 8.28 \\
MC-CSM with LBT \cite{taherian_LBT_2022} & 13.2 & 14.8 & 3.33 & 0.910 & 9.62 \\
TFGridNet \cite{wang_TFGridNetIntegratingFull_2022} & 22.8 & 24.9 & 4.08 & 0.980 & 6.76 \\
\hline
SpatialNet-small (prop.)  & 21.3 & 23.2 & 3.99 & 0.974 & 7.05 \\ 
SpatialNet-large (prop.)  & \textbf{25.1} & \textbf{27.1} & \textbf{4.17} & \textbf{0.986} & \textbf{6.70} \\ 
\hline\hline
\end{tabular}
}
\vspace{-0.2cm}
\end{table}

\subsection{Results on WHAMR!}
WHAMR! \cite{maciejewski_WHAMRNoisyReverberant_2020} extends the WSJ0-2mix dataset \cite{hershey_DeepClusteringDiscriminative_2016} by adding noise recorded with binaural microphones in urban environments and introducing reverberation to the speech sources.
The SNR is randomly sampled from -6 to +3 dB.
We test on the same version of the dataset as in \cite{wang_TFGridNetIntegratingFull_2022}, namely the 8 kHz and 'min' version.
The first channel is taken as the reference channel. Table \ref{table_whamr} shows the results. The proposed SpatialNet-large slightly outperforms TFGridNet. 
This WHAMR! dataset is more difficult to process than SMS-WSJ, as it involves severe environmental noise, and all the performance scores of unprocessed signals are lower. Thence, it is more challenging for the proposed network to process by mainly exploiting spatial information, especially when only two microphones are provided. By contrast, TFGridNet adopts a strong network to fully exploit spectral information.  

\begin{table}[tbp]
\setlength\tabcolsep{4pt}
\renewcommand{\arraystretch}{1.3}
\caption{Results on 2-channel WHAMR! dataset. 
$^{\star}$ 
denote that the scores are quoted from \cite{wang_TFGridNetIntegratingFull_2022}.
}
\label{table_whamr}
\centering
\resizebox{\linewidth}{!}{
\begin{tabular}{lcccc}
\hline\hline
\textbf{Method} & SISDR (dB) & SDR (dB) & NB-PESQ & eSTOI \\
\hline
unproc. & -6.1 & -3.5 & 1.41 & 0.317 \\
\hline
Multi-TasNet $^{\star}$ \cite{zhang_mc_convtasnet_2020} & 6.0 &  &  &  \\
\makecell[l]{Multi-TasNet with\\ \ \ speaker extraction} $^{\star}$ \cite{zhang_TimeDomainSpeechExtraction_2021} & 7.3 &  &  &  \\
TFGridNet \cite{wang_TFGridNetIntegratingFull_2022} & 13.7 & 14.8 & \textbf{3.16} & 0.868 \\
\hline
SpatialNet-small (prop.)  & 11.8 & 13.1 & 2.93 & 0.826 \\ 
SpatialNet-large (prop.)  & \textbf{14.1} & \textbf{15.0} & \textbf{3.16} & \textbf{0.870} \\ 
\hline\hline
\end{tabular}
}
\vspace{-0.2cm}
\end{table}

\subsection{Results on Spatialized WSJ0-2mix}
The Spatialized WSJ0-2mix dataset \cite{wang_multi-channel_2018} is a spatialized extension of the WSJ0-2mix dataset \cite{hershey_DeepClusteringDiscriminative_2016}.
The clean speech signals in WSJ0-2mix are convolved with 8-channel simulated RIRs.
The microphone array geometry is randomly sampled with an aperture size drawn from 15 cm to 25 cm.
T60 is randomly drawn from 0.2 s to 0.6 s.
The speech pairs are overlapped in ``max" or ``min" type  \cite{hershey_DeepClusteringDiscriminative_2016, wang_multi-channel_2018}.
The relative energy ratio between speakers uniformly distributes in $[-5,\ 5]$ dB.
The sampling rate could be either 8 kHz or 16 kHz.
For fair comparison with other works, we use the first four channels and take the first channel reverberant image as the target signal, namely only speech separation is conducted.

\begin{table}[t]
\setlength\tabcolsep{2pt}
\renewcommand{\arraystretch}{1.2}
\caption{Results on the 4-channel Spatialized WSJ0-2mix dataset. $^{\star}$, $^{\ast}$ and $^{\dagger}$ denote that the scores are quoted from \cite{wang_multi-channel_2018}, \cite{ochiai_beam-tasnet_2020} and \cite{chen_BeamGuidedTasNetIterative_2022}, respectively. 
}
\label{table_random_array}
\centering
\resizebox{\linewidth}{!}{
\begin{tabular}{lcccc}
\toprule
\multirow{2}{*}{\textbf{Method}} & \multicolumn{2}{c}{8 kHz} & \multicolumn{2}{c}{16 kHz}\\
~ & NB-PESQ & SDR (dB) & WB-PESQ & SDR (dB) \\
\midrule
\multicolumn{5}{c}{\textbf{Speech signals are overlapped in the ``max" type}} \\
unproc. & 2.00 & 0.1 & 1.45 & 0.1 \\
\hline
MC Deep Clustering \cite{wang_multi-channel_2018} & - & 9.4$^{\star}$ & - & - \\
MC-TasNet \cite{gu_EndtoEndMultiChannelSpeech_2019}  & - & - & - & 12.1$^{\ast}$ \\
Beam-TasNet \cite{ochiai_beam-tasnet_2020} & - & 17.4$^{\dagger}$ & - & 16.8$^{\ast}$ \\
\makecell[l]{Beam-Guided TasNet\ (iter=4)}
 \cite{chen_BeamGuidedTasNetIterative_2022} & - & 21.5$^{\dagger}$ & - & - \\
Oracle signal MVDR & - & 23.5$^{\dagger}$ & - & 21.7$^{\ast}$ \\
\hline
SpatialNet-Small (prop.) & 4.35 & 27.6 & 4.33 & 28.0 \\
SpatialNet-Large (prop.) & \textbf{4.43} & \textbf{32.7} & \textbf{4.46} & \textbf{32.4} \\
\hline \hline 
\multicolumn{5}{c}{\textbf{Speech signals are overlapped in the ``min" type}} \\
unproc. & 1.81 & 0.2 & 1.29 & 0.0 \\
\hline
DasFormer \cite{wang_DasformerDeepAlternating_2023} & 4.33 & 26.1 & - & - \\
SpatialNet-small (prop.) & 4.35 & 26.8 & 4.34 & 26.6 \\
SpatialNet-large (prop.) & \textbf{4.43} & \textbf{31.7} & \textbf{4.45} & \textbf{31.2} \\
\bottomrule
\end{tabular}
}
\vspace{-0.2cm}
\end{table}

Table \ref{table_random_array} shows the speech separation results.
We can see that the neural beamformers, i.e. Beam-TasNet and Beam-Guided TasNet, show the superiority of beamforming over the binary-mask-based method (i.e. MC Deep Clustering), and the time-domain end-to-end method (i.e. MC-TasNet).
DasFormer obtains very good separation results, by alternating frame-wise (full-band) and band-wise (narrow-band) self-attention networks. 
The proposed SpatialNet-small outperforms all the comparison methods. Different from DasFormer that uses the same self-attention scheme for both full-band and narrow-band learning, the proposed SpatialNet uses a heavy narrow-band conformer block and a light convolutional-linear cross-band block to more efficiently exploit the spatial information. Moreover, SpatialNet-large achieves almost perfect speech quality. 

In this dataset, the geometry of microphone array is varying, which is more challenging for the end-to-end networks compared to the beamforming-based methods as beamforming is array-agnostic, while the end-to-end networks needs to generalize across different arrays. Even though, DasFormer and the proposed SpatialNet still achieve good performance, which shows their capability of array-generalization.

\subsection{Results on LibriCSS}
LibriCSS \cite{chen_ContinuousSpeechSeparation_2020} is a meeting-like dataset recorded in a regular meeting room by playing utterances sampled from LibriSpeech \cite{panayotov_LibrispeechASRCorpus_2015} with loudspeakers.
There are 10 sessions in LibriCSS, among which \textit{session0} can be used for hyper-parameter tuning.
Different speech overlap ratios are set, including 0\% with short inter-utterance silence (0S) or long inter-utterance silence (0L), 10\%, 20\%, 30\% and 40\%.
A 7-channel circular array is used.
An ASR system is provided along with this dataset to measure the speech enhancement performance.
We report the performance of our networks for utterance-wise evaluation, where utterances are split with oracle boundary information of each utterance.  

\begin{table}[tpb]
\setlength\tabcolsep{3pt}
\renewcommand{\arraystretch}{1.2}
\caption{WER (\%) for LibriCSS (7-channel, utterance evaluation). 
}
\label{table_libricss_results}
\centering
\begin{tabular}{lcccccc}
\hline\hline
\multirow{2}{*}{\textbf{Method}} & \multicolumn{6}{c}{Overlap ratio (\%)}\\
~ & 0S & 0L & 10 & 20 & 30 & 40 \\
\hline
unproc. & 11.8 & 11.7 & 18.8 & 27.2 & 35.6 & 43.3 \\
\hline
LSTM $+$ MVDR \cite{chen_ContinuousSpeechSeparation_2020} & 8.4 & 8.3 & 11.6 & 15.8 & 18.7 & 21.7 \\
Conformer-base $+$ MVDR \cite{chen_continuous_2021} & 7.3 & 7.3 & 9.6 & 11.9 & 13.9 & 15.9 \\
Conformer-large $+$ MVDR \cite{chen_continuous_2021} & 7.2 & 7.5 & 9.6 & 11.3 & 13.7 & 15.1 \\
MISO$_1$-BF-MISO$_3$ \cite{wang_MultimicrophoneComplexSpectral_2021} & 5.8 & 5.8 & 5.9 & 6.5 & 7.7 & 8.3 \\
\hline
SpatialNet-small (prop.) & \textbf{5.5} & 5.5 & 5.8 & 6.7 & 7.8 & 8.6 \\ 
SpatialNet-large (prop.) & 5.6 & \textbf{5.4} & \textbf{5.6} & \textbf{5.9} & \textbf{6.5} & \textbf{6.9} \\ 
\hline\hline
\end{tabular}
\vspace{-0.2cm}
\end{table}

As this dataset only provides the evaluation data, we use simulated signals to train our networks.
The simulated array has the same geometry with the LibriCSS array.
According to the overlap mode of evaluation data, two speech streams are simulated and mixed. One stream is the speech signal of one (target) speaker. The other stream could include the speech signal of zero, one or two (non-overlapped) speakers, with the proportion of about 33\%, 53\% and 14\%, respectively.    
When the other stream includes one speaker, the speech overlap ratio is randomly sampled in [10\%, 100\%].
When the other stream includes two speakers, the two speakers talk respectively in the beginning and ending parts of the stream, with a 0.1 $\sim$ 1.0 s silence in between. 
The clean speech signals are sampled from \textit{train-clean-100} and \textit{train-clean-360} sets of LibriSpeech.
Multichannel diffuse noises are generated using the toolkit \cite{habets2008generating} with the single-channel ambient noise signals of Reverb Challenge dataset \cite{kinoshita_ReverbChallengeCommon_2013}.
SNR between reverberant mixture and noise is randomly sampled in [5, 20] dB.
T60 is randomly sampled in [0.2, 1.0] s.
The gpuRIR toolkit \cite{diaz-guerra_gpurir_2021} is used for RIR generation.
To train the network with one possible empty output stream, the negative of source-aggregated SDR (SA-SDR) \cite{vonneumann_SASDRNovelLoss_2022} is used as the loss function. 
The WER of \textit{session0} is taken as the validation metric of network training. To produce a stable ASR performance, the network weights of the last ten checkpoints/epochs (relative to the best epoch) are averaged. 
At inference, the evaluation utterances are first chunked to 4-second segments and processed by the network, with 2-second overlapping between consecutive segments.
The final output is formed by stitching the segment-level outputs according to the similarity of overlapped parts.

The proposed networks are compared with three neural beamformers \cite{chen_ContinuousSpeechSeparation_2020, chen_continuous_2021, wang_MultimicrophoneComplexSpectral_2021}. 
Table \ref{table_libricss_results} shows the ASR performance.
WERs can be greatly reduced by the neural beamformers no matter for single-speaker or multi-speaker situations. The proposed SpatialNet-large largely outperforms other methods, which demonstrates the effectiveness of proposed network as an ASR front-end technique on real data.

\subsection{Results on Reverb Challenge Dataset}
The Reverb Challenge dataset \cite{kinoshita_ReverbChallengeCommon_2013} includes simulated data (SimData) and real-recorded data (RealData), for evaluating speech dereverberation performance. A 8-channel circular array with a diameter of 20 cm is used. The sampling rate is 16 kHz. Two different speaker-to-array distances (near=50 cm and far=200 cm) are evaluated. We use the official evaluation metrics of Reverb Challenge.
Readers can refer to \cite{kinoshita_ReverbChallengeCommon_2013} for more details about the dataset and evaluation metrics.

The Reverb Challenge training set only consists of 24 real-recorded RIRs, which are insufficient for training the multichannel speech enhancement networks. 
For that reason, we use 40,000 simulated RIRs with the same array geometry to train our networks.
The direct-path signal is taken as the training and prediction target.
Accordingly, the direct-path signal is taken as the reference signal for intrusive evaluation metrics. However, the Reverb Challenge traditionally takes the dry (source) speech as the reference signal. For fair comparison, both dry speech and direct-path signal will be tested.  
To align with \cite{wang_DeepLearningBased_2020}, 2.5 ms around the peak value of measured RIRs are considered as the direct-path component.
ASR performance is evaluated using the best pretrained ASR checkpoint \footnote{https://github.com/espnet/espnet/tree/master/egs2/reverb/asr1\#transformer-asr--transformer--lm---speedperturbation--specaug--applying-rir-and-noise-data-on-the-fly} for this dataset in ESPnet \cite{li2020espnet}. 

We compare with the WPE- and beamforming-based methods, including WPE, WPE$+$BeamformIt and WPD, and the deep learning based methods proposed in \cite{wang_DeepLearningBased_2020}. The ESPnet implementation of WPE and WPE$+$BeamformIt are used. For WPD, the implementation \footnote{\url{https://github.com/Emrys365/espnet/blob/wsj1_mix_spatialized/espnet/nets/pytorch_backend/frontends/WPD_beamfomer_v6.py}} is used, where the predication delay and the number of taps are set to 3 and 5, respectively.

Table \ref{table_reverb} shows the enhancement results. 
The proposed SpatialNet outperforms the comparison methods in terms of CD, FWSSNR, PESQ by a large margin, which demonstrates that the proposed network is able to largely suppress reverberation and meanwhile maintain the speech quality. Table \ref{table_reverb_asr} shows the ASR results. On SimData, all methods achieve good performance, as their WERs are close to the ones of clean speech. On RealData, the proposed method performs much better than WPE- and beamforming-based methods, especially for the even difficult ``far'' data. These results show that the proposed SpatialNet is very efficient for dereverberation in terms of both speech enhancement and ASR. Note that, the proposed networks are trained using simulated data, and thus suffer from the simulation-to-real generalization problem more or less. If the networks can be trained with sufficient real-recorded data, the performance can be further improved. 

\begin{table}[t]
\setlength\tabcolsep{2pt}
\renewcommand{\arraystretch}{1.3}
\caption{Enhancement Results on 8-channel Reverb dataset. 
PESQ with $^*$ stands for "NB PESQ MOS", otherwise for "WB MOS LQO".}
\label{table_reverb}
\centering
\resizebox{\linewidth}{!}{
\begin{tabular}{lcccccc}
\hline\hline
\multirow{2}{*}{\textbf{Method}} & \multicolumn{5}{c}{SimData} & \multicolumn{1}{c}{RealData}\\
~ & CD$\downarrow$ & LLR$\downarrow$ & FWSSNR$\uparrow$ & PESQ$\uparrow$  & SRMR$\uparrow$ & SRMR$\uparrow$ \\
\hline
\multicolumn{7}{c}{Dry source signal is taken as the reference signal} \\
unproc. & 3.97 & 0.58 & 3.62 & 1.50 & 3.68 & 3.18 \\
\hline
WPE \cite{nakatani_SpeechDereverberationBased_2010,drude_NARAWPEPythonPackage_2018} & 3.48 & 0.45 & 6.05 & 1.96 & 4.85 & 5.04 \\
WPE$+$BeamformIt \cite{li2020espnet} & 2.66 & \textbf{0.34} & 8.72 & 2.22 & 5.17 & 5.62\\
WPD \cite{nakatani_UnifiedConvolutionalBeamformer_2019,zhang_EndtoEndFarFieldSpeech_2020} & 2.89 & 0.54 & 5.94 & 1.87 & \textbf{5.98} & 5.72 \\
\hline
SpatialNet-small (prop.) & 1.52 & 0.44 & 12.51 & 3.66 & 4.65 & \textbf{7.18} \\
SpatialNet-large (prop.) & \textbf{1.36} & 0.39 & \textbf{13.04} & \textbf{3.83} & 4.43 & 6.88 \\
\hline\hline
\multicolumn{7}{c}{Direct-path signal is taken as the reference signal} \\
unproc. & 5.04 & 0.67 & 8.38 & 2.37$^*$ &  &  \\\hline
WPE \cite{nakatani_SpeechDereverberationBased_2010,drude_NARAWPEPythonPackage_2018} & 4.75 & 0.53 & 11.42 & 2.83$^*$ &  & \\
WPE$+$BeamformIt \cite{li2020espnet} & 3.94 & 0.49 & 12.52 & 3.12$^*$ &  &  \\
WPD \cite{nakatani_UnifiedConvolutionalBeamformer_2019,zhang_EndtoEndFarFieldSpeech_2020} & 4.15 & 0.49 & 10.50 & 2.80$^*$ &  &  \\
Wang et al. \cite{wang_DeepLearningBased_2020} & 2.78 & 0.39 & 18.75 & 3.71$^*$ &  &  \\
\hline
SpatialNet-small (prop.) & 2.32 & 0.32 & 19.41 & 3.93$^*$ &  &  \\
SpatialNet-large (prop.) & \textbf{2.26} & \textbf{0.29} & \textbf{21.80} & \textbf{4.05}$^*$ &  &  \\
\hline\hline
\end{tabular}
}
\vspace{-0.2cm}
\end{table}

\begin{table}[t]
\setlength\tabcolsep{8pt}
\renewcommand{\arraystretch}{1.2}
\caption{WER (\%) results on 8-channel Reverb dataset.}
\label{table_reverb_asr}
\centering
\begin{tabular}{l|cc|cc}
\hline\hline
\multirow{2}{*}{\textbf{Method}} & \multicolumn{2}{c|}{SimData} & \multicolumn{2}{c}{RealData}\\
~ & far & near & far & near \\
\hline
unproc. & 4.9 & 3.7 & 6.5 & 5.9 \\
clean & 3.4 & 3.4 & - & - \\
\hline
WPE \cite{nakatani_SpeechDereverberationBased_2010,drude_NARAWPEPythonPackage_2018} & 3.8 & \textbf{3.5} & 4.6 & 4.9 \\
WPE$+$BeamformIt \cite{li2020espnet} & 3.7 & \textbf{3.5} & 4.4 & 3.6 \\
WPD \cite{nakatani_UnifiedConvolutionalBeamformer_2019,zhang_EndtoEndFarFieldSpeech_2020} & 3.8 & \textbf{3.5} & 4.5 & 4.1 \\
\hline
SpatialNet-small (prop.)  & \textbf{3.6} & 3.7 & \textbf{2.8} & 3.4 \\
SpatialNet-large (prop.)  & \textbf{3.6} & 3.6 & 3.1 & \textbf{3.2} \\
\hline\hline
\end{tabular}
\vspace{-0.2cm}
\end{table}

\subsection{Results on CHiME 3/4 Challenge Dataset}
The CHiME 3/4 challenge \cite{barker_ThirdCHiMESpeech_2015} provides both simulated and real-recorded datasets. Speech utterances from WSJ0 corpus are used. 
Multichannel speech are recorded with a speaker holding a tablet device equipped with 6 microphones.
\addnote[ref chn 2]{1}{The fifth channel is taken as the reference channel following \cite{yang2022mcnet,wang_ComplexSpectralMapping_2020,shimada_UnsupervisedSpeechEnhancement_2019,tolooshams_ChannelAttentionDenseUNet_2020}.}
Multichannel background noise are recorded in four real environments, including bus, cafe, pedestrian area, and street junction. The sampling rate is 16 kHz.
In the literature, the ASR performance on this dataset have been extensively developed, and reached a very high performance level. Therefore, we only report the speech enhancement performance on the simulated dataset. 
Reverberation in this dataset is very small, as the multichannel speech is either simulated by delaying a single-channel speech, or recorded in a booth. Thence, this dataset only performs speech denoising, with real-recorded multichannel noise. 

Table \ref{table_chime4_sim} shows the enhancement performance.
We can see that, narrow-band deep filtering (NBDF) \cite{li_narrow-band_2019} with a simple two-layer BLSTM network obtains better performance than MNMF beamforming \cite{shimada_UnsupervisedSpeechEnhancement_2019} and time-domain end-to-end FasNet$+$TAC \cite{luo_end--end_2020}, which proves the efficiency of learning narrow-band information. 
By flipping the first LSTM layer of NBDF from time axis to the frequency axis, FT-JNF \cite{tesch_InsightsDeepNonLinear_2023} jointly learns narrow-band and cross-band information, and further improves the enhancement performance. This demonstrates that the cross-band information is complementary to narrow-band information.
Based on NBDF and FT-JNF, McNet \cite{yang2022mcnet} uses additional LSTM layers to learn the spectral difference between speech and noise, and further largely improve the performance.
\addnote[chime wang]{1}{\cite{wang_ComplexSpectralMapping_2020} is a two-stage neural beamformer with two temporal convolutional networks and two MVDR beamformers, and it prominently outperforms the aforementioned methods, which demonstrates the advancement of neural beamforming in low-reverberation scenario.}

The proposed SpatialNet shares a similar spirit with FT-JNF but is equipped with more powerful neural networks for respectively learning the narrow-band and cross-band spatial information. 
The results show that, SpatialNet achieves comparable results with \cite{wang_ComplexSpectralMapping_2020}, and largely outperforms other comparison methods, which shows the effectiveness of the proposed network architecture. 
It is worth noting that SpatialNet-small slightly outperforms SpatialNet-large on this dataset, which is different from the results on other datasets. 
This is possibly because of the low-reverberation of this dataset, for which less information are needed to learn than other datasets.

\begin{table}[tpb]
\setlength\tabcolsep{2pt}
\renewcommand{\arraystretch}{1.2}
\caption{Speech enhancement results on 6-channel simulated CHiME 3/4 dataset. $^{\star}$ denote that the scores are quoted from \cite{yang2022mcnet}, respectively. }
\label{table_chime4_sim}
\centering
\resizebox{\linewidth}{!}{
\begin{tabular}{lccccc}
\hline\hline
\textbf{Method} & NB-PESQ & WB-PESQ & STOI & SI-SDR (dB) & SDR (dB) \\
\hline
unproc. & 1.82 & 1.27 & 0.870 & 7.5 & 7.5 \\
\hline
MNMF Beamforming \cite{shimada_UnsupervisedSpeechEnhancement_2019} & - & - & 0.940 & - & 16.2 \\
FasNet$+$TAC $^{\star}$ \cite{luo_end--end_2020} & 2.53 & 1.91 & 0.944 & 16.0 & 16.6 \\
CA Dense U-net  \cite{tolooshams_ChannelAttentionDenseUNet_2020} & - & 2.44 & - & - & 18.6 \\
NBDF $^{\star}$ \cite{li_narrow-band_2019} & 2.74 & 2.13 & 0.950 & 15.7 & 16.6 \\
FT-JNF $^{\star}$ \cite{tesch_InsightsDeepNonLinear_2023} & 3.20 & 2.61 & 0.967 & 17.6 & 18.3 \\
McNet \cite{yang2022mcnet} & 3.38 & 2.73 & 0.976 & 19.2 & 19.6 \\
Wang et al. \cite{wang_ComplexSpectralMapping_2020} & - & - & \textbf{0.986} & 22.0 & \textbf{22.4} \\
Oracle MVDR & 2.49 & 1.94 & 0.970 & 17.3 & 17.7 \\
\hline
SpatialNet-small (prop.) & \textbf{3.49} & 2.88 & 0.983 & \textbf{22.1} & 22.3 \\ 
SpatialNet-large (prop.) & 3.45 & \textbf{2.89} & 0.982 & 21.8 & 22.1 \\
\hline
\hline\hline
\end{tabular}
}
\vspace{-0.0cm}
\end{table}

\subsection{Summary of Computational Complexity and Model Size}
\addnote[flops and size]{1}{
In this section, we summarize the number of floating point operations (FLOPs) and the number of parameters (\#Param) of the proposed SpatialNet and those comparison methods (involved in this paper) either open-sourced code or reported relevant quantities in their papers. 
For the methods with code available, we measure the FLOPs and \#Param for both 8 kHz and 16 kHz data using 6-channel, 4-second, 2-speaker utterances. FLOPs is computed in the way described in \footref{flops}.  
The number of channels and the number of speakers may differ from the one reported in their original papers, but their contributions to FLOPs and \#Param are negligible. Except that, FLOPs of FaSNet+TAC \cite{luo_end--end_2020} linearly increases with the increase of the number of microphone channels, thus we report both the 2-channel and 6-channel results for FaSNet+TAC.
NBDF \cite{li_narrow-band_2019}, FT-JNF \cite{tesch_InsightsDeepNonLinear_2023}, McNet \cite{yang2022mcnet}, DasFormer \cite{wang_DasformerDeepAlternating_2023}, TFGridNet \cite{wang_TFGridNetIntegratingFull_2022} and the proposed SpatialNet are STFT-domain methods, for which STFT window length and hop size are respectively set to 32 ms and 16 ms according to their original papers except that TFGridNet is 32 ms and 8 ms according to its paper. TFGridNet stacks two networks (DNN$_1$ and DNN$_2$) with the same architecture (having a minor difference for the input dimension), thus we report twice the FLOPs and \#Param of one DNN. We use the code provided in \footref{tfgridnet code}, but the DNN is configured according to the configuration for the SMS-WSJ dataset reported in \cite{wang_TFGridNetIntegratingFull_2022}.
Similarly, twice \#Param is reported for the method presented in \cite{wang_ComplexSpectralMapping_2020}. 
FaSNet+TAC \cite{luo_end--end_2020} and Beam Guided TasNet \cite{chen_BeamGuidedTasNetIterative_2022} are time-domain methods, and the window length and hop size are respectively 4ms and 2 ms in FaSNet+TAC, and 2ms and 1ms in Beam Guided TasNet.

Table \ref{table_flops} shows the results. The STFT-domain methods, i.e. NBDF, FT-JNF, McNet, DasFormer, TFGridNet and the proposed SpatialNet, all perform narrow-band processing, namely processing frequencies independently, which leads to a relative larger computational complexity, compared to the full-band methods, such as FaSNet+TAC. In addition, the narrow-band processing methods will double the FLOPs when the sampling rate is increased from 8 kHz to 16 kHz. On the other hand, the model size of narrow-band processing methods are normally small, as a small network can  process the information of one frequency well, and the network parameters are shared by all frequencies. The proposed SpatialNet-small has a relatively small FLOPs and model size among the narrow-band processing methods, and achieve outstanding speech enhancement performance as presented previously. TFGridNet and the proposed SpatialNet-large have much larger FLOPs than other methods, and correspondingly achieve much better performance on most of tasks than other methods.  
}

\begin{table}[tbp]
\setlength\tabcolsep{3pt}
\renewcommand{\arraystretch}{1.3}
\caption{Computational complexity and model size of the proposed network and comparison methods.}
\label{table_flops}
\centering
\resizebox{1\linewidth}{!}{
\begin{tabular}{ccccc}
\hline\hline
\multirow{3}{*}{\textbf{Method}} & \multicolumn{2}{c}{\textbf{8 kHz}} & \multicolumn{2}{c}{\textbf{16 kHz}} \\
 & \makecell[c]{FLOPs\\(G/s)} & \makecell[c]{\#Param\\(M)} & \makecell[c]{FLOPs\\(G/s)} & \makecell[c]{\#Param\\(M)}\\
\hline
NBDF \tablefootnote{https://github.com/Audio-WestlakeU/Narrowband\_DeepFiltering} \cite{li_narrow-band_2019} & 19.5 & 1.2 & 38.9 & 1.2 \\
FT-JNF \tablefootnote{The same architecture with NBDF, but flips the first LSTM to frequency axis.} \cite{tesch_InsightsDeepNonLinear_2023} & 19.5 & 1.2 & 38.9 & 1.2 \\
McNet \tablefootnote{https://github.com/Audio-WestlakeU/McNet} \cite{yang2022mcnet} & 29.7 & 1.9 & 59.2 & 1.9 \\
\multirow{2}{*}{\makecell[c]{FaSNet+TAC \tablefootnote{https://github.com/yluo42/TAC\label{tac-code}} \cite{luo_end--end_2020}}} & 9.7 & 2.7 & 10.1 & 2.8 \\
  & 26.7 & 2.7 & 27.9 & 2.8 \\
\makecell[c]{Beam Guided TasNet \tablefootnote{https://github.com/hangtingchen/Beam-Guided-TasNet\label{bg code}} (iter=2) \cite{chen_BeamGuidedTasNetIterative_2022}} & 34.4 & 5.9 & 85.8 & 6.1 \\
DasFormer \tablefootnote{the source code is kindly provided by its author} \cite{wang_DasformerDeepAlternating_2023} & 33.3 & 2.2 & 76.4 & 2.2 \\
TFGridNet \tablefootnote{https://github.com/espnet/espnet/blob/master/espnet2/enh/separator/ tfgridnet\_separator.py\label{tfgridnet code}} \cite{wang_TFGridNetIntegratingFull_2022} & 348.4 & 11.0 & 695.6 & 11.2 \\
Wang et al. \cite{wang_ComplexSpectralMapping_2020} & - & - & - & 26.0 \\
\hline
SpatialNet-small (prop.) & 23.1 & 1.2 & 46.3 & 1.6 \\
SpatialNet-large (prop.) & 119.0 & 6.5 & 237.9 & 7.3 \\
\hline\hline
\end{tabular}
}
\vspace{-0.2cm}
\end{table}

\section{Conclusions}
In this paper, we propose the SpatialNet to extensively leverage spatial information for multichannel joint speech separation, denoising and dereverberation.
SpatialNet is mainly composed of interleaved narrow-band and cross-band blocks, to respectively exploit narrow-band and cross-band spatial information. 
Experiments show that the proposed SpatialNet outperforms other state-of-the-art methods on various simulated and real-world tasks. The excellent performance of the proposed SpatialNet verifies that spatial information is highly discriminative for speech separation, denoising and dereverberation, and SpatialNet is effective to fully leverage these information. In addition, the proposed SpatialNet suffers little from the spectral generalization problem, and thus preforms well on a cross-language speech enhancement task. Currently, the proposed SpatialNet is only designed for offline (non-causal) processing, and the online version will be developed in the future.  

\bibliographystyle{IEEEtran}

\bibliography{mybib}

\end{document}